\documentclass[preprint,aps,showpacs,showkeys,preprintnumbers,amsmath,amssymb,amsfonts]{revtex4}
\usepackage{graphics}% Include figure files
\usepackage{graphicx}% Include figure files
\usepackage{epsfig}
\usepackage{dcolumn}% Align table columns on decimal point
\usepackage{bm}% bold math
\usepackage{latexsym}% To use symbols in LATEX2.09 which removed by 2e

\begin{document}

\title{Thermal Resonance Fusion}

\author{ Bao-Guo \,Dong }
\affiliation{\\
Department of Nuclear Physics, China Institute of Atomic Energy,
P.O. Box 275 (10), Beijing 102413, China }
\date{\today}

\begin{abstract}
We first show a possible mechanism to create a new type of nuclear fusion, thermal resonance fusion, i.e.
low energy nuclear fusion with thermal resonance of light nuclei or atoms, such as deuterium or tritium. The fusion of two light nuclei has to overcome the Coulomb barrier between these two nuclei to reach up to the interacting region of nuclear force. We found nuclear fusion could be realized with thermal vibrations of crystal lattice atoms coupling with light atoms at low energy by resonance to overcome this Coulomb barrier. Thermal resonances combining with tunnel effects can greatly enhance the probability of the deuterium fusion to the detectable level. Our low energy nuclear fusion mechanism research - thermal resonance fusion mechanism results demonstrate how these light nuclei or atoms, such as deuterium, can be fused in the crystal of metal, such as Ni or alloy, with synthetic thermal vibrations
and resonances at different modes and energies experimentally. The probability of tunnel effect at different resonance  energy given by the WKB method is shown that indicates the thermal resonance fusion mode, especially combined with the tunnel effect, is possible and feasible. But the penetrating probability decreases very sharply when the input resonance energy decreases less than 3 keV, so for thermal resonance fusion, the key point is to increase the resonance peak or make the resonance sharp enough to the acceptable energy level by the suitable compound catalysts, and it is better to reach up more than 3 keV to make the penetrating probability larger than $10^{-10}$.
\end{abstract}

%crystal lattice atom thermal vibration, interacting boundary of nuclear force}

%\pacs{25.60.Pj, 24.30.Gd, 63.90.+t, 33.40.+f, 89.30.Jj }

\maketitle

%One Sentence Summary: A new type of fusion, thermal resonance fusion,
%i. e. low energy nuclear fusion with thermal resonance of light atoms, such as deuterium or tritium are predicted, and a
%new research field between nuclear physics and atomic physics is
%opened.

Thermonuclear fusions of light nuclei or atoms in stars in nature is well known and in hydrogen bombs as well as in laboratories, such as deuterium (D) or tritium (T), i.e. D+D or D+T, was performed experimentally and predicted theoretically early \cite{NF15}.

\begin{figure}
\includegraphics*[width=8.6cm]{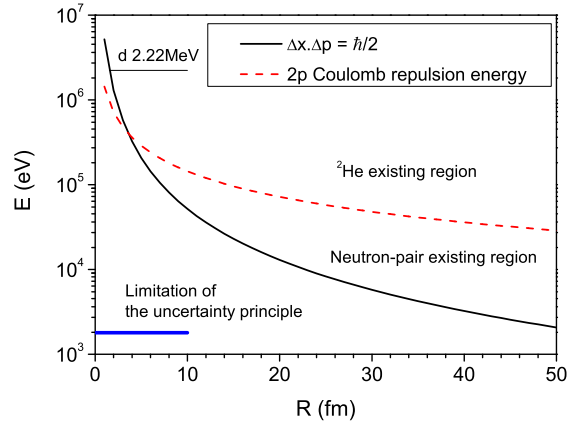}
\caption{\label{trf} (Color online) Coulomb barrier between two protons or deuterons. And limitation of the uncertainty principle, $\Delta x \cdot \Delta p \geq \hbar/2$, to the neutron-pair formed for nuclear force. The estimated interacting region of nuclear force is shown in blue line, about 10 fm. Above
the limit line is the possible binding state region and below the
non-binding region. The two-nucleons binding energy required as a function of the possible binding state region, including Coulomb repulsive energy of proton+proton or D+D or D+T, is shown.}
\end{figure}

　　Experimental results indicate that nickel catalyzing or adsorbing D gas can release heat more than the quantity expected \cite{CF14}. It is disputed that this experimental phenomenon of heat production is a chemical process  or low energy nuclear fusion process at present. This redundant heat comes from what  procedure and mechanism and is worth for exploratory investigation.　Actually, low energy D fusion catalyzed by nickel can be investigated and confirmed experimentally. Because D fusion must overcome Coulomb barrier between these two deuterons to reach the interacting region of nuclear force due to its interacting short-distance compared with Coulomb force, see Fig. \ref{trf}, general theory suggests the fusion mechanisms include:
%{\it  Super high temperature fusion} This thermonuclear fusion overcomes Coulomb barrier with kinetic energy of light nuclei, Such as D+T fusion in hydrogen bombs.
%{\it  Quantum tunnel effect fusion}  The other possible mechanism of low energy nuclear fusion is through the  quantum tunnel effect to realize the fusion, such as D+D. But it should be a very low probability process.

\begin{description}
\item[ Super high temperature fusion] This thermonuclear fusion overcomes Coulomb barrier with kinetic energy of light nuclei, Such as D+T fusion in hydrogen bombs.

\item[ Quantum tunnel effect fusion]  The other possible mechanism of low energy nuclear fusion is through the  quantum tunnel effect to realize the fusion, such as D+D. But it should be a very low probability process.
\end{description}　　

These two nuclear fusion mechanisms and their combination could not well interpret the phenomena of the above low energy experimental observation of heat produced \cite{CF14}.

We first suggest and show a sort of possible mechanism to create a new type of nuclear fusion, thermal resonance fusion, i.e.
low energy nuclear fusion with thermal resonance of light nuclei or atoms, such as D or T, through the head-on collision mechanism of the high density coupling thermal vibration and resonance of light nuclei in crystal realizes the fusion.
We found nuclear fusion could be realized with thermal vibrations of crystal atoms coupling with light atoms at low energy by resonance to overcome this Coulomb barrier. Our low energy D fusion mechanism research - thermal resonance fusion mechanism results demonstrate how these
light nuclei or atoms can be fused in the crystal of metal, such as Ni or alloy, with synthetic thermal vibrations
and resonances and their couplings at different modes and energies experimentally.

\begin{figure}
\includegraphics*[width=8.6cm]{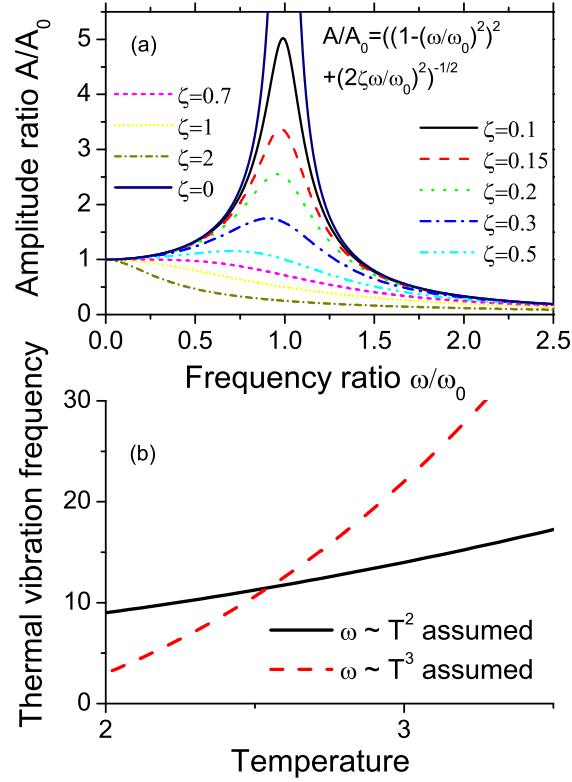}
\caption{\label{trfgb} (Color online) (a) Total vibrating relative amplitudes enhanced by the two resonance deuterons to fusion. To overcome the Coulomb barrier with resonance energy and tunnel effect, the enhanced effect of relative amplitudes of two resonance deuterons comes from different damping parameters, $\zeta$ is shown in $A/A_0$ as a function of $\omega/\omega_o$, $A/A_0=((1-(\omega/\omega_o)^2)^2+(2\zeta\omega/\omega_o)^2)^{-1/2}$.
(b) Dependence of thermal vibration frequency $\omega(T)$ on temperature $T$. Different atoms to form the alloy would have different $\omega \propto \omega(T)$ curves and different resonance $Q$ values. The resonance occurs at the cross of two assumed curves. Sketch for understanding the fusion mechanism directly and clearly.}
\end{figure}

By vibration theory \cite{Fa98}, we can simplify these thermal vibrations as simple harmonic oscillators, see Fig. \ref{trfgb}(a),
with differential equation of motion
$\stackrel{..}{x} + 2\zeta\omega_0 \stackrel{.}{x} + \omega^2_0 x = A_0 \omega^2_0 cos\omega t$,
where $\omega^2_0=k/m$, $m$, $k$, $\omega_0$, $A$, $\omega$, $x$, and $\zeta$ is the mass, spring constant, natural frequency, amplitude, input frequency, displacement from the equilibrium position, and damping parameter, respectively. $A_0=F_0/k$, where $F_0$ is the amplitude of input simple harmonic driving force.
And we can express their motion as, $x=Acos(\omega t+\phi)$ and their couplings as,  $x=x_1+x_2=A_1cos(\omega_1 t+\phi_1)+ A_2cos(\omega_2 t+\phi_2)$, where $\phi$ and $t$ are the phase and time. The total vibrating energy, $E$, of harmonic vibrations is $E=kA^2/2$. The maximum resultant amplitude of two harmonic vibrations is $A_1 + A_2$ in the same direction for both the same and different $\omega$.
The quality factor or Q factor $Q = 1/(2\beta) = \sqrt{mk}/d$, where $\beta$ is the damping ratio, and d is the damping coefficient, defined by the equation $F_{d} = -dV$, where $V$ is the velocity and $F_{d}$ is the damping force.  Then when the system resonates, i.e. $\omega=\omega_0$, we have $E=\alpha Q^2/2 = kF^2_0 Q^2 /(2\omega^2_0)$ with a parameter $\alpha = kF^2_0 /\omega^2_0$.

　　
Deuterium gas adsorbed by nickel makes its density enhance. Nickel atoms, i.e. nickel nuclei, in solid alloy thermally vibrate at crystal lattice, and such thermal vibration would make deuterons do analogous forced vibration and coupling with their natural thermal vibrations to cause the resonance of two D atoms or deuterons under suitable catalyzed conditions. The opponent motion of two neighbour deuterons, based on the assistance of resonance interaction mechanism, such as the multiple thermal resonance, possibly would make the instantaneous velocity of opponent thermal vibration of two neighbour deuterons, namely transient energy, reach up to the degree or order to overcome Coulomb barrier between them, thereby to cause the thermal resonance fusion of the two neighbour deuterons, i.e. realize low energy nuclear fusion by resonance of thermal vibration.

In solid state physics \cite{FJX80}, vibrations of crystal lattice are very complicated issues. Structure of most metals at normal solid state is crystal, such as nickel Ni, palladium Pd, or platinum Pt, with each atom at its crystal lattice to do thermal vibrations, forming different moving modes. These small amplitude thermal vibrations include collective vibrations, such as optical wave and phonon wave modes, and coupling vibrations can be treated as linear vibrator with frequency $\nu_0=1/(2\pi)\sqrt{(c_0/m)}$, here $c_0$ is resilience constant and $m$ the mass. The energy of atoms vibrating around the equilibrium crystal lattice is related to its temperature and by classical theory, each freedom of atoms with mean energy of $K_BT$, $K_B$ is Boltzmann constant and $T$ the temperature. So the atom mean energy in each freedom is about 0.1 eV at 1000 K for example. By quantum theory, the energy of crystal lattices is quantized and with frequency $\omega$ it is
%\begin{equation}
$E_n(\omega)=(n+1/2)\hbar\omega$,
%\end{equation}
and with Boltzmann distribution function as a function of $T$,
%\begin{equation}
$f_B(T,\omega)=e^{-n\hbar\omega/(K_BT)}$.
%\end{equation}
This means atomic vibration energy having distribution, especially the vibrations of atoms absorbed, and some atoms may have much higher energy than the rest, for example to reach up to 10 eV or higher. This indicates that there could be atoms with distributing energy high enough to cause the fusion or at least enhance the fusion probability of D+D especially coupling with the tunnel effect.

The D+D or D+T fusion is performed with nuclear force. The nucleon-nucleon interaction or nuclear force is not full clear to understand by now due to its complicacy and is dependent on
relative distance, relative movement situation, relative momentum of
interacting nucleons, spin, isospin, energy, density etc. There
exists no derivation of the nucleon-nucleon force from first
principles. So it is difficult to confirm the features of nuclear
force by the existing theoretical nucleon-nucleon interactions, such
as the bare nucleon-nucleon forces, microscopic effective
interactions, and phenomenological effective interactions,
especially under the extreme conditions at several eV low energy
and near the interacting boundary of nuclear force. The main
features of nuclear force are short distance and saturation character.
By the model theories supported by experiments, when two nucleons
are near and interact with strong interaction, the main attractive
effect comes from the s-wave, and p-wave for repulsive. For two
different nucleons, obey the Pauli principle and total wave function
asymmetry under the two particle exchange, the spin direction can be
parallel or antiparallel, and the attractive interaction strength of
parallel is even larger than the antiparallel. For two same
nucleons, fermion and identical particle, only the spin antiparallel
state can exist the relative movement s-state, which decreases the
probability in the s-state and weakens the interaction \cite{Hu87}.
For these the experimental evidence is deuterium, $^2$H, exists and
no $^2$He observed in nature.

In general, if an exchange of the coordinates causes the change of
the space coordinate part of wave function is symmetry there is an
attractive force between the two nucleons, and if asymmetry the
force is repulsive. The long distance part of nuclear force is attractive and
the short part is repulsive, which is confirmed and demonstrated by
experiments \cite{Hu87}.
%The properties of nuclear force
%are unclear by now, especially in the interacting boundary region,
%which could be investigated experimentally in this way.
%Experimenters can accurately test features of low energy DD
%collisions first for easy detectability in experiments.

For D+D the interacting situation is more complicated due to its including two nucleons. But we can approximately consider D as a vibrating point particle when the distance of two deuterons is not in the interacting region of nuclear force.
There are many vibration modes or methods to couple them together for different $\alpha$ and $Q$ to produce resonance with much higher energy.

If we intend to use the new effective and suitable method, the thermal resonance fusion technology, to perform and realize the low energy nuclear fusion of two deuterons and produce the heat, we can select
appropriate density, and absorbing temperature, metal, such as Ni, Pd, or Pt, or special alloy as
the catalyst for the sharp peak resonance, i.e. to get larger $\alpha$ and $Q$. Then we can try to carry out the low energy nuclear fusion experimentally.

%To increase D density,
%we have to develop suitable method and technology to increase D gas density and make the interval between two deuterons shorter and shorter.
%one possible method is the multi-grade high pressure compression method, which may be adopted to increase the D gas density.
%This high density storage technology could be applied to hydrogen gas storage in industry and daily usage.

\begin{figure}
\includegraphics*[width=8.6cm]{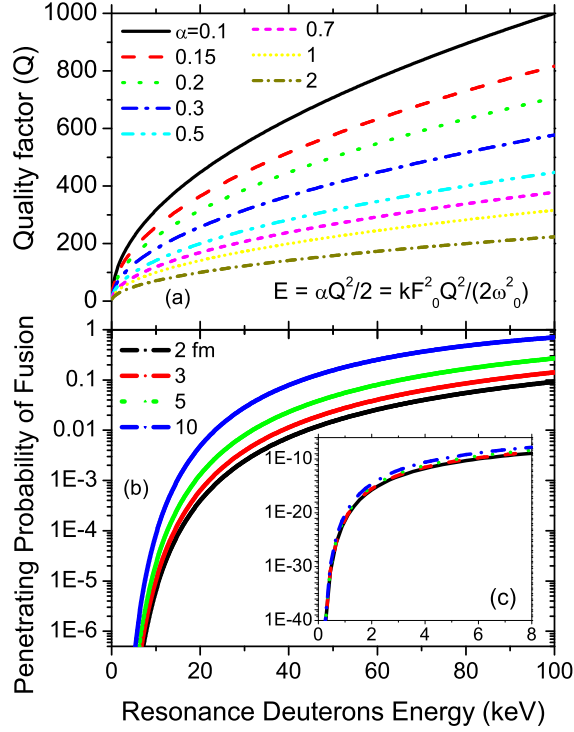}
\caption{\label{trf3} (Color online) (a) Resonance energy reached by different interacting strength of or with $\alpha$ and quality factor, Q. (b) Penetrating probability of two resonance deuterons to fusion as a function of input energy, given by the WKB method \cite{Ze81}. To overcome the Coulomb barrier with resonance energy and tunnel effect, the reached interval between two resonance deuterons is 2, 3, 5, and 10 fm, respectively, up to the nuclear force region. (c) The same as (b) but for energy less than 8 keV.  The penetrating probability decreases very sharply with input energy decreasing less than 3 keV.}
\end{figure}

 The extremely difficult challenge to fuse
the deuteron pairs with resonance is experimentally to create the suitable resonance mode really to overcome the Coulomb repulsive force between deuterons to perform the fusion.
%We have to seek and find the proper vibration modes,
%method and mechanism from the sophisticated catalyst crystal lattice thermal vibrations to form the resonance deuteron pairs,  make high density D gas fuse, namely low energy nuclear fusion. One possible mechanism or way to form the deuteron
%pairs is by the resonances between two deuterons to reach up near enough to the
%interacting boundary of nuclear force.
The other mechanism is the thermal resonance combining with the tunnel effect to realize the fusion, see Fig. \ref{trf3}.

%We can obtain resonance with energy on the order of eV from thermal vibrations directly by the traditional methods, such as vibration mechanics  \cite{Fa98}.
For example, if the quality factor $Q $ of the coupling system is $Q = 500$, when $\alpha = 0.1$ it resonances the maximum resonance energy $E_m$ is $E_m=25$ keV, and the maximum amplitude $A/A_0=1/(2\zeta)$ could become $10^{3}$ with $\zeta=0.0005$, see Figs. \ref{trfgb}(a) and \ref{trf3}.

On the other hand, the resonance makes the deuteron energy increase that at least would enhance the probability of two deuterons to fusion based on and coupling with the tunnel effect. Under the above condition, i.e. 25 keV, the probability of the tunnel effect reached up to 0.0011, 0.0017, 0.0036, and 0.0132 for the interval of 2, 3,5, and 10 fm, respectively.

If 1 g D-gas is absorbed by Ni and it releases the heat of 1 J, it needs about the D number of 1 J/4.04MeV = $3\times 10^{12}$ to fusion for D + D $\rightarrow$ T + p, for example. The fusion probability of two deuterons of 1 g D-gas is $3\times 10^{12} / (1/4\times6.023\times10^{23}) = 2\times10^{-11}$.

This probability is very low requirement in the D resonance energy comparing with the tunnel effect probability. But the penetrating probability decreases very sharply when the input resonance energy decreases less than 3 keV, see Fig. \ref{trf3} (b) and (c). So the key point for the fusion is to increase the D resonance energy to the acceptable energy level, more than 3 keV to make the penetrating probability larger than $10^{-10}$, for example.

%\begin{figure}
%\includegraphics*[width=8.6cm]{trftem.eps}
%\caption{\label{trf3} (Color online) Penetrating probability of two resonance deuterons to fusion as a function of input energy for the very low input energy, given by the WKB method. To overcome the Coulomb barrier with resonance energy and tunnel effect, the reached interval between two resonance deuterons is 2, 3, 5, and 10 fm, respectively, up to the nuclear force region. The penetrating probability decreases very sharply with input energy decreasing less than 3 keV.}
%\end{figure}

The challenges include that one has to make the low D gas density
high enough to enhance the interaction probability to form the resonance deuteron pairs, create suitable catalyst
and explore or detect the boundary of nuclear force in deuteron by both theory and experiment. For example, if two absorbed D atoms thermally vibrate with $\omega_D \propto \omega_D(T)$, and the abutted on crystal lattice atoms in alloy thermally vibrate with $\omega_C \propto \omega_C(T)$, both increase with increasing temperature $T$ but with different slope, the resonance will occur at the cross of two curves, see Fig. \ref{trfgb}(b). Different alloy would have different $\omega \propto \omega(T)$ curves and different $Q$ values. To perform the measurable thermal resonance fusion, it needs to find out suitable alloy with high $Q$ value experimentally, and investigate in physics, technology and corresponding equipments.

%\begin{figure}
%\includegraphics*[width=8.6cm]{trf5.eps}
%\caption{\label{trf5} (Color online) Dependence of thermal vibration frequency $\omega(T)$ on temperature $T$. Different atoms to form the alloy would have different $\omega \propto \omega(T)$ curves and different resonance $Q$ values. The resonance occurs at the cross of two assumed curves. Sketch for understanding the fusion mechanism directly and clearly.}
%\end{figure}

Ultimately, these deuteron pairs fuse to form T or $^3$He or $^4$He by the common nuclear fusion mode
and to release heat. The detecting methods of the fusion processes and the existence of the neutrons are difficult for experiments.
To determine the coupling sophisticated crystal lattice thermal vibrations and these dynamical processes is an open
problem. The possible characteristic neutron, proton or photon emitted by the two deuteron fusion
 could be as the detected signal for the fusion.

The Coulomb repulsive energy of two deuterons is shown in Fig. \ref{trf}. The limitation of first principles has been considered carefully.
The limitation of the uncertainty principle, $\Delta x \cdot \Delta
p \geq \hbar/2$, to the pair formed is shown in Fig. \ref{trf}. This
principle gives the possible distance between the two interacting
neutrons and the minimum binding energy they have to need if they
form a pair with spin zero. The binding energy of D is
2.2246 MeV so the nearest distance for the two nucleons is 1.526 fm
under the limitation of the uncertainty principle. When the distance
is 10 fm the binding energy needed for two neutrons decreases to 52 keV, see Fig.
\ref{trf}. It would be easy to determine exactly the interacting boundary of nuclear force with neutrons due to no Coulomb force. And this also provides a very good field to investigate deeply the features of nuclear force which is now understood as a residual effect of the strong interaction force in particle physics, i.e. residual strong force, especially in the boundary region, the competitive feature between nuclear force and Coulomb force in the same region, and the contradiction between the uncertainty principle and the asymptotic freedom and where their effectively and suitably working boundary is.

Once the thermal resonance fusion, i.e.
low energy nuclear fusion with thermal resonance of light nuclei or atoms, such as D or T is performed and confirmed experimentally, it can be as a new type of clean and safe energy source.

In summary, we have suggested a possible mechanism to create a new type of nuclear fusion, thermal resonance fusion, i.e. low energy nuclear fusion with thermal resonance of light nuclei or atoms, such as D or T. The fusion of two light nuclei has to overcome the Coulomb barrier between them to reach the interacting region of nuclear force. We found low energy nuclear fusion could be realized with thermal vibrations of crystal lattice atoms coupling
with light atoms at low energy by resonance to overcome this Coulomb barrier. The other mechanism is the thermal resonance combining with the tunnel effect to realize the fusion. Our low energy D fusion mechanism research - thermal resonance fusion mechanism results demonstrate how these
light nuclei or atoms can be fused in the crystals, such as Ni or alloy, with the sophisticated synthetic thermal vibrations and resonances at different modes and energies experimentally.
But the penetrating probability decreases very sharply with input resonance energy decreasing less than 3 keV, so for thermal resonance fusion, the key point is to increase the D resonance peak, or make the resonance sharp enough to the acceptable energy level by the suitable compound catalysts, for example, more than 3 keV to make the penetrating probability larger than $10^{-10}$.

\end{document}